\documentclass[12pt]{iopart}
\usepackage{iopams}  
\usepackage[dvips]{graphicx,color}
\usepackage{hyperref}  
\newcommand{\qs}{Q_{\mathrm{s}}}

\newcommand{\nc}{N_\mathrm{c}}
\newcommand{\cf}{C_\mathrm{F}}
\newcommand{\as}{\alpha_{\mathrm{s}}}

\newcommand{\ud}{\mathrm{d}}
\newcommand{\xt}{\mathbf{x}_T}

\newcommand{\kt}{{\mathbf{k}_T}}

\newcommand{\gev}{\textrm{ GeV}}

\begin{document}

\title{
The glasma initial state at the LHC
}

\author{T. Lappi}

\address{
Institut de Physique Th\'eorique\\
CEA/DSM/SPhT, CEA/Saclay\\
F-91191 Gif-sur-Yvette Cedex, France
}
\ead{tuomas.lappi@cea.fr}
\begin{abstract}
We present results from numerical Classical Yang Mills calculations of the initial
stage Glasma field configurations in a relativistic heavy ion collision.
We compute the initial gluon multiplicity from RHIC to  LHC energies.
The initial conditions for the classical field computation are taken from a dipole model 
parametrization tested on HERA data, meaning that
all the parameters, including the normalization and the  value of the saturation scale,
are fixed from DIS data.
\end{abstract}



\section{Introduction: the Glasma}
At high energy  the small $x$ part of the wavefunction of a hadron or nucleus
can be understood as a classical color field radiated
by static color sources formed by the large $x$ degrees of  freedom~\cite{McLerran:1994ni}.
This description, known as the color glass condensate (for reviews see e.g.
\cite{Iancu:2003xm,Weigert:2005us}), provides
a common framework for understanding both small $x$ deep inelastic scattering (DIS)
and the initial stages of relativistic heavy ion collisions.
The cross section for small $x$ DIS can be expressed in terms of the correlator
of two Wilson lines (i.e. the dipole cross section), and the initial condition
for the classical fields that dominate the first fraction of a fermi of a
heavy ion collision is determined by these same Wilson lines.

The earliest stage of an ultrarelativistic
heavy ion collision is a coherent, classical field configuration of two colliding 
sheets of Color Glass. This first fraction of a fermi
of the collision, in transition towards a equilibrated
quark gluon plasma, is what we refer to as the Glasma \cite{Lappi:2006fp,Lappi:2006hq}.
The color fields of the two nuclei are transverse electric and magnetic
fields on the light cone.
The glasma fields left over in the region between the two nuclei
after the collision at times $0 < \tau \lesssim 1/\qs$
are, however, longitudinal along the beam axis.
They fields depend on the transverse
coordinate on a length scale of order $1/\qs$.
As the system expands the fields are diluted and can
be treated as particles. This lowest order contribution, in the MV model, 
has been computed numerically  by Classical Yang-Mills (CYM)
computations of gluon production  in heavy ion 
collisions~\cite{Krasnitz:1998ns,Krasnitz:2001qu,Lappi:2003bi}.

\begin{figure}
\includegraphics[width=0.44\textwidth]{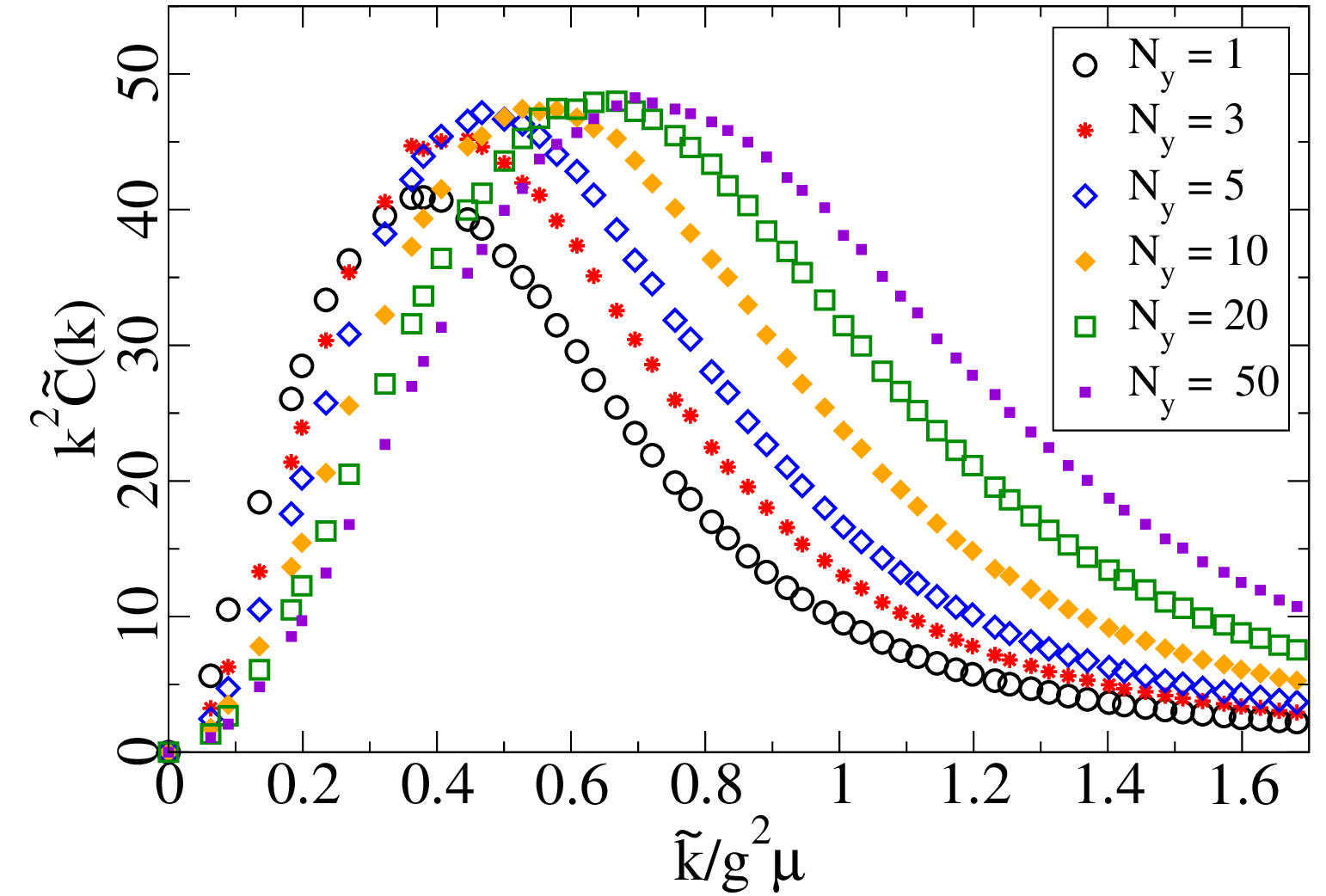}
\hfill
\includegraphics[width=0.44\textwidth]{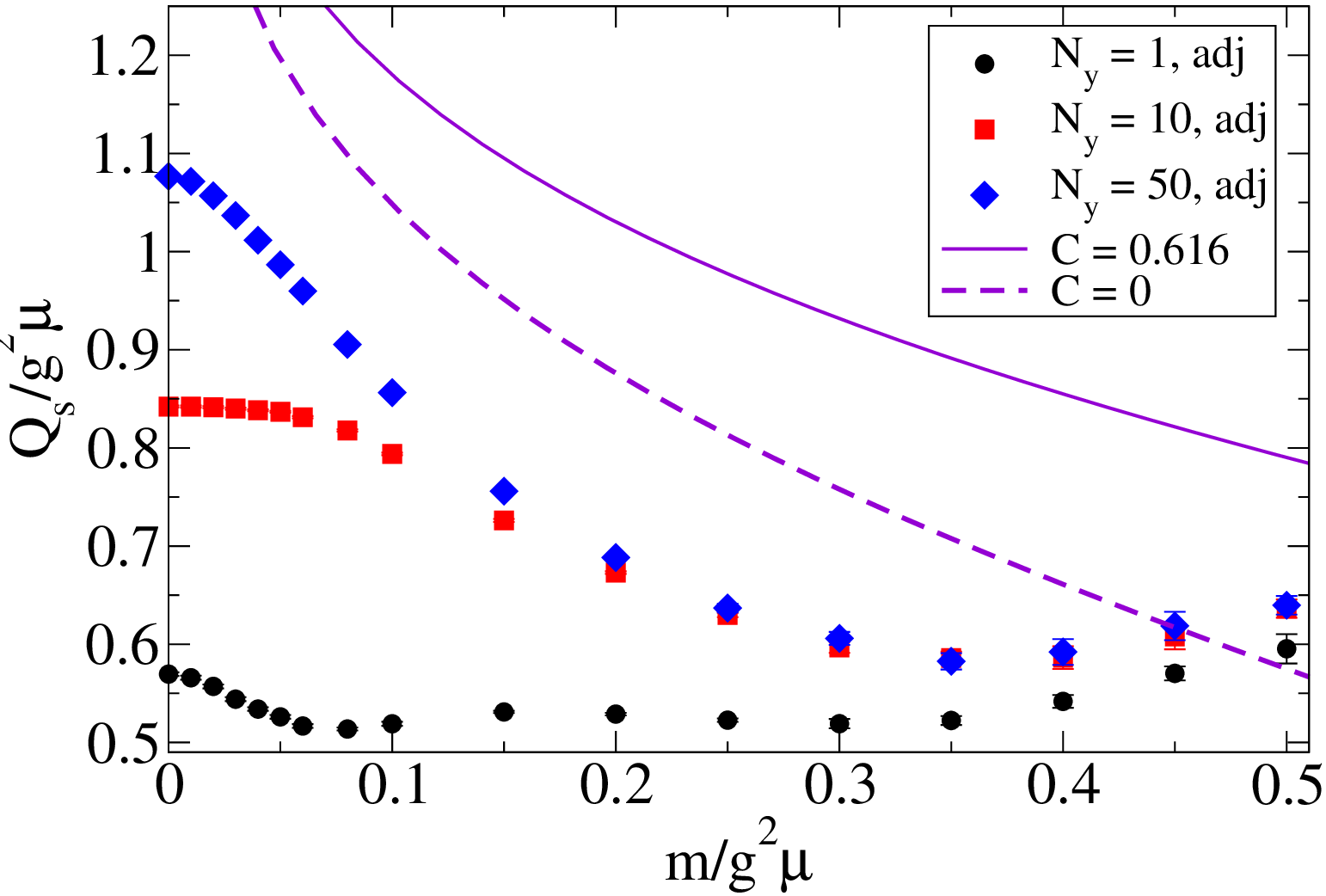}
\caption{
Left: The Wilson line correlator vs. $\kt/g^2\mu$ in the MV model
with different discretizations of the longitudinal coordinate. The value
$N_y = 1$ corresponds to the implementation in the CYM calculations.
Right: The saturation scale (in the adjoint representation) 
in units of $g^2\mu$ for different values of the infrared
regulator $m$ used in the calculation. The straight and dashed lines correspond to 
the analytical ($N_y \to \infty$) estimate
$\frac{\qs^2 }{\left( g^2 \mu \right)^2} = 
\frac{\nc}{2 \pi } \left[ \ln \frac{g^2 \mu}{m} + C \right]$
with different values for the constant $C$. The value corresponding
to the CYM calculations is $\qs = 0.57 g^2\mu$
(at $m=0$ and $N_y=1$).
}
\label{fig:qs}
\end{figure}

In order to relate calculations of the initial gluon multiplicity to final state 
observables at RHIC and LHC, some simplifying assumptions have to be made. We will 
in this simple study assume that the initial gluon multiplicity is approximately 
equal to that of the final state hadrons (charged and neutral); about
1000/1150 per unit rapidity at RHIC  with $\sqrt{s} = 130 A \gev / 200 A \gev$.
This assumption can be justified by either rapid thermalization of the system and
entropy conservation during its hydrodynamical evolution, or the phenomenological 
assumption of parton--hadron duality between the gluon and hadron multiplicities.
Note that for the relation between the initial and final energy densities 
these two assumptions lead to very different interpretations.

\section{Value of the saturation scale}

To make genuine predictions for the multiplicities in heavy ion collisions
we must be able to obtain the value of the saturation scale from DIS measurements at
HERA. For the purposes of this study we will do this using
 two dipole cross section parametrisations,  where the
impact parameter dependence is modeled and constrained using diffractive HERA data. 
These are the IPsat (or Kowalski-Teaney) and the bCGC parametrisations. 
The IPSat model \cite{Kowalski:2003hm,Kowalski:2007rw}, is built from
an eikonalised DGLAP-evolved gluon distribution function. Thus it contains the right
color transparency limit at large $Q^2$. 
The bCGC model \cite{Kowalski:2006hc} is an impact parameter dependent
version of the IIM~\cite{Iancu:2003ge} parametrization, which includes 
the main effects of BK evolution. For RHIC energies the values
of the saturation scale at median $b$ from these parametrizations are
are $\qs \approx 1.1\gev$ (IPsat) and $\qs \approx 1.0 \gev$ (bCGC)
\cite{Kowalski:2007rw}.

So far the CYM computations of the Glasma initial state have been 
performed with initial conditions from the MV 
model~\cite{McLerran:1994ni}, where the color charge density
is determined by a phenomenological parameter $g^2\mu$. 
In order to use DIS data to set the value for $g^2\mu$, one 
must consistently determine the value of the Wilson line correlation length 
(the saturation scale) in implementation of the MV model used in the numerical 
computations. This comparison has only recently been performed in 
Ref.~\cite{Lappi:2007ku} (see also the related discussion in 
Ref.~\cite{Fukushima:2007ki}). The results are summarized in Fig.~\ref{fig:qs}.
The obtained value $\qs/g^2\mu \approx 0.57$ means that the HERA-extracted values quoted 
above are in remarkable agreement with the earlier estimate $g^2\mu \approx 2 \gev$
based on RHIC phenomenology alone~\cite{Krasnitz:2001qu,Lappi:2003bi}.

\begin{figure}
\includegraphics[width=0.32\textwidth]{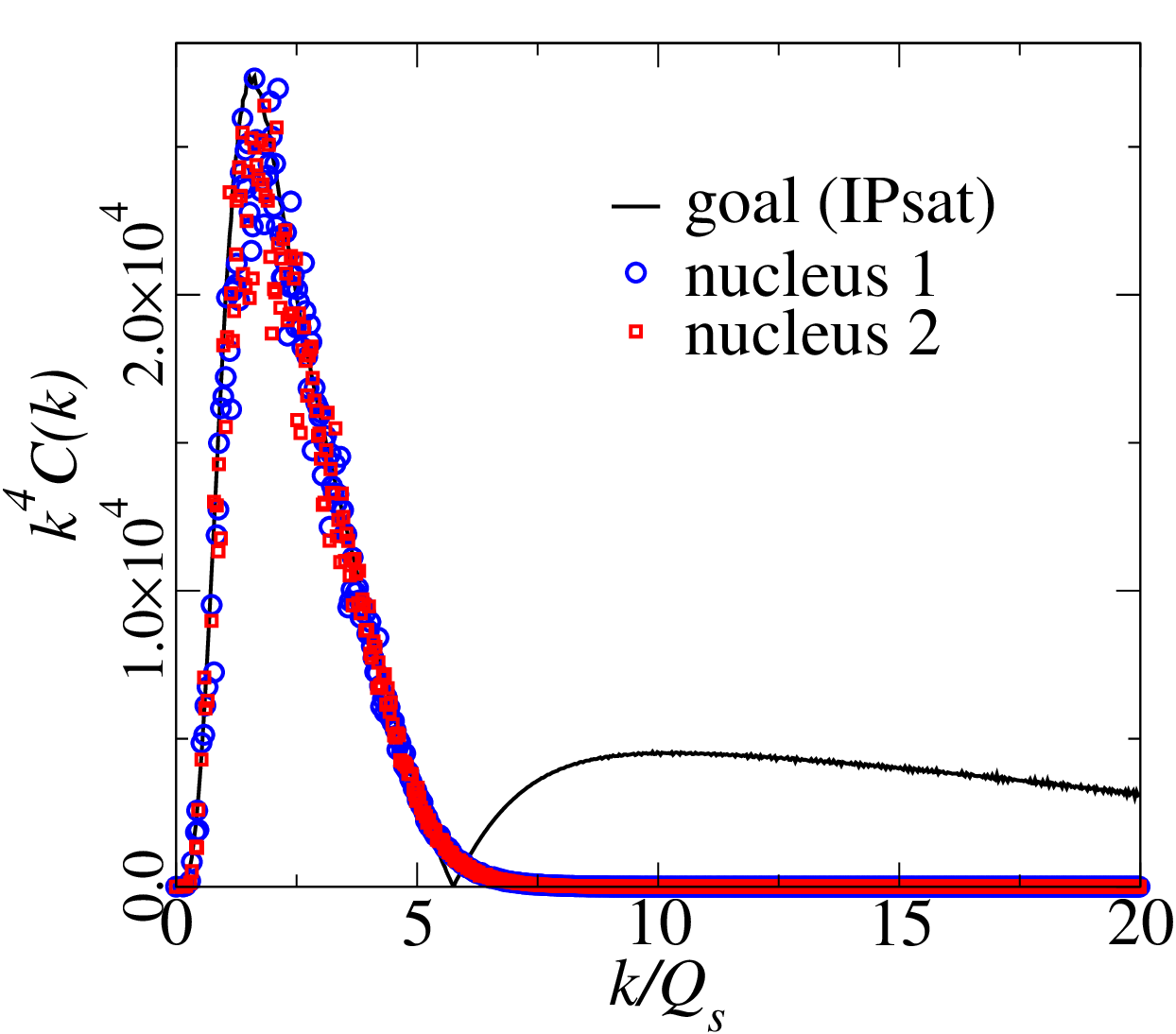}
\hfill
\includegraphics[width=0.32\textwidth]{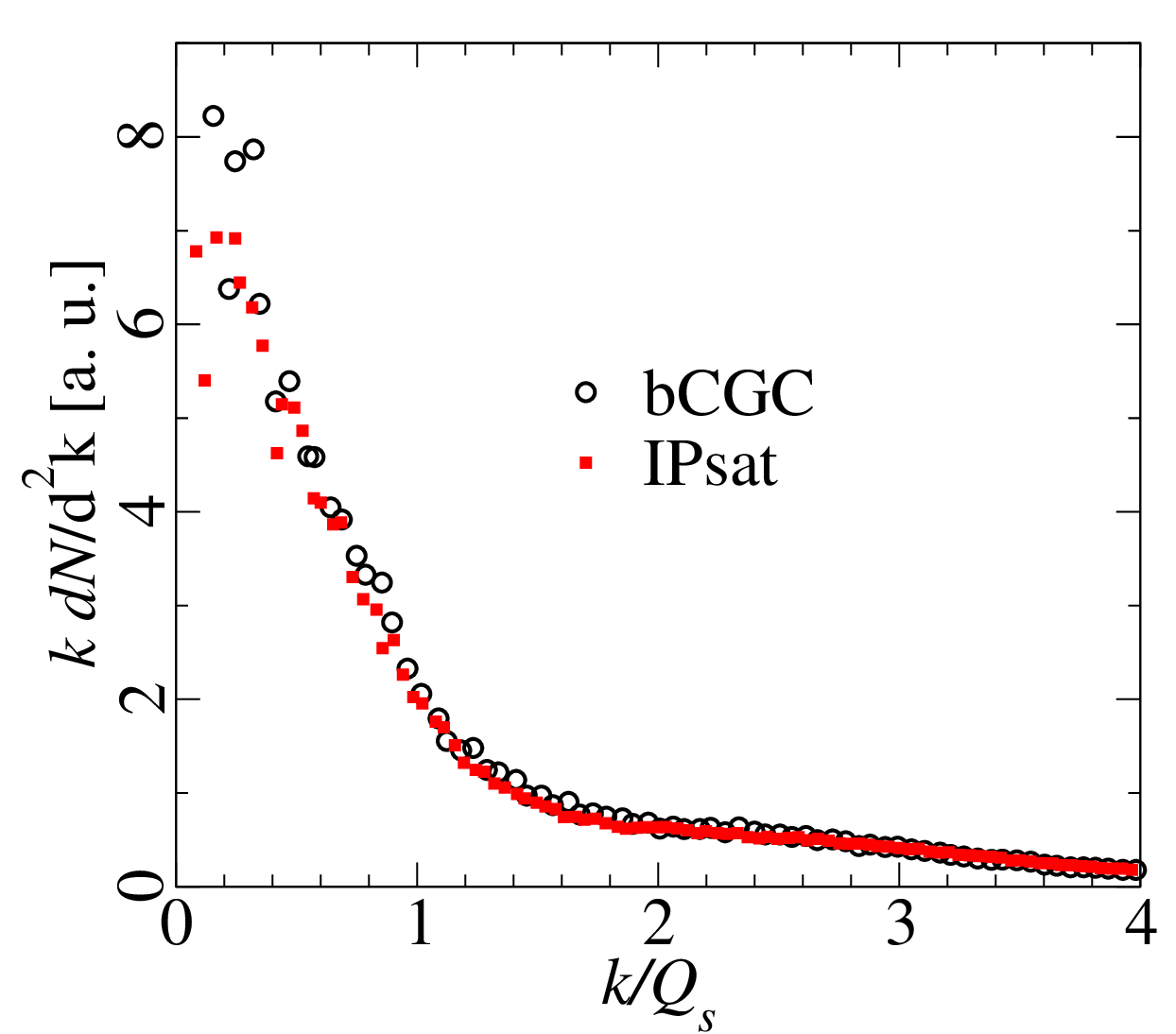}
\hfill
\includegraphics[width=0.32\textwidth]{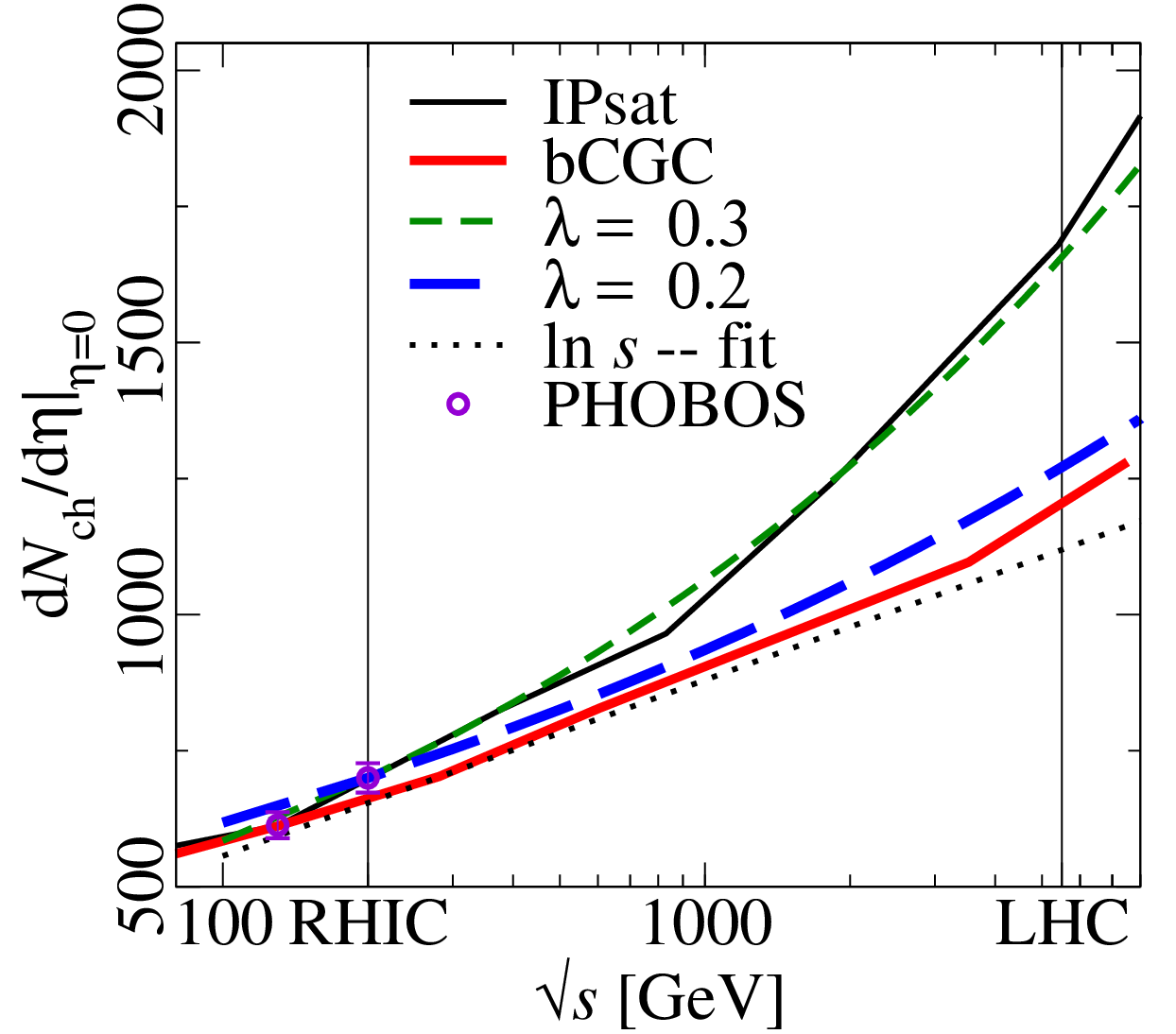}
\caption{
Left: Momentum space Wilson line correlator (scaled by $\kt^4$) 
in the IPsat model at RHIC energies
(``goal'') and the same correlator from two field configurations constructed 
to reproduce it, on average.
Center: Spectrum of produced gluons at LHC energy using IPsat and bCGC dipole 
cross sections.
Right: Energy dependence of charged particle multiplicity from the IPsat and bCGC
models, from the MV model assuming $\qs^2 \sim x^{-\lambda}$ and from 
a $A + B \ln s$-fit to lower energy data.
}
\label{fig:dipxs}

\end{figure}

\section{Beyond the MV model, LHC predictions}

It is possible to go beyond the MV model in the CYM calculation of the 
glasma fields. This can be done by supplementing the given Wilson line correlator
with additional assumptions of the statistics of the large $x$-sources.
We adopt the commonly used ``nonlocal Gaussian'' approximation, in which 
the color fields $A^+(\xt,x^-)$ in the Wilson line 
$U = \mathrm{P}\exp\{i g \int \ud x^- A^+ \}$ at different values of the
longitudinal coordinate $x^-$ are taken to be independent and have a Gaussian 
distribution. With this additional assumption one can numerically construct 
the Wilson lines; the result of this procedure is shown in 
Fig.~\ref{fig:dipxs} (left).  At large momenta the $\kt$-space correlator is negative 
(absolute value plotted); only the positive part is used. 
This leads to a problematic behavior of the gluon spectrum at large $\kt$ 
(see Fig.~\ref{fig:dipxs} center), but
for bulk quantities, such as the total multiplicity,
the effect is not significant.

\begin{table}
\begin{center}
\begin{tabular}{|rl|r|r|r|r|r|}
\hline
\hline
$\sqrt{s}/A$ & model  &$c_x$& $x$ & $\qs \, [\gev] $ & $\ud N_g/\ud \eta$ & c \\
\hline
\hline
200&GeV   IPsat  & 1.00 & $5.56\times 10^{-3}$ & $1.1 $ & 1000 & 1.2\\
200&GeV bCGC   & 1.00 & $5.05\times 10^{-3}$ & $1.0 $ & 850 & 1.2 \\
\hline
5500&GeV IPsat & 0.58 & $1.98 \times 10^{-4}$ & $1.9 $ & 2900 & 1.1 \\
5500&GeV bCGC  & 0.24 & $0.67 \times 10^{-4}$ & $1.6 $ & 2200 & 1.1 \\
\hline
\end{tabular}
\end{center}
\caption{Summary of CYM results. Calculated on a $512^2$-lattice at 
$\tau = 1.6/\qs$.} \label{tab:results}
\end{table}

It is convenient to parametrize our results in terms of the ``gluon liberation 
coefficient'' $c$~\cite{Mueller:1999fp} which is defined using the ratio of the
gluon multiplicity to $\qs^2$ by
$
\frac{\ud N}{\ud^2 \xt \ud y} = c \frac{\cf \qs^2}{2 \pi^2 \as}.
$
The MV model calculation leads~\cite{Lappi:2007ku} to the result $c\approx 1.1$.
We find that both the IPsat and bCGC models lead to very similar values,
within 10\% of the MV model result, for the liberation coefficient $c$.
The gluon spectrum obtained using the IPsat or bCGC parametrisations is
shown in Fig.~\ref{fig:dipxs} (center). 
There is a  remaining uncertainty in the value of $x$ used to determine 
the saturation scale in an ion-ion collision where, unlike in DIS, it is not 
precisely determined by the collision kinematics. Parametrically one should 
take $x = c_x \qs / \sqrt{s}$ with $c_x$ of order one. Taking 
$c_x = 1$  yields a lightly smaller gluon multiplicity than our assumption
based on the measured RHIC charged multiplicity. In making a prediction for LHC 
we adjust the constant $c_x$ in order to obtain  $\ud N_g/\ud \eta = 1150$
at RHIC 200~GeV. Our
first preliminary results for RHIC and LHC energies from this study are 
summarized in  Table~\ref{tab:results}

The $\sqrt{s}$ dependence of the charged particle multiplicity from RHIC to
 LHC energies is shown in Fig.~\ref{fig:dipxs} (right). There is a significant
difference between the bCGC and IPsat models, resulting from the much weaker 
energy dependence of $\qs$ in the former. Thus the LHC measurements of particle 
multiplicities will significantly constrain our understanding of the high energy 
evolution in the Color Glass Condensate. In the meanwhile, to make LHC prediction
more accurate, further theoretical and phenomenological 
 work is needed to understand effects such as the impact
parameter dependence of the dipole cross section and running coupling and pomeron 
loop effects on the evolution.

\ack

The author thanks R. Venugopalan for discussions. 

\section*{References}

\bibliographystyle{JHEP-2mod}
\bibliography{spires}

\end{document}